\pdfoutput=1
\documentclass[12pt,a4paper]{article}

\usepackage{ifthen} 
\newboolean{pdflatex}
\setboolean{pdflatex}{true} 

\newboolean{articletitles}
\setboolean{articletitles}{true} 

\newboolean{uprightparticles}
\setboolean{uprightparticles}{false} 

\def\paperauthors{LHCb collaboration} 
\def\paperasciititle{Search for B(s)*0 -> mu+ mu- in Bc+ -> pi+ mu+ mu- decays} 
\def\papertitle{Search for $B_{(s)}^{*0}\to\mu^+\mu^-$ in $B_c^+\to\pi^+\mu^+\mu^-$ decays}
\def\paperkeywords{{High Energy Physics}, {LHCb}} 
\def\papercopyright{\the\year\ CERN for the benefit of the LHCb collaboration} 
\def\paperlicence{CC BY 4.0 licence}
\def\paperlicenceurl{https://creativecommons.org/licenses/by/4.0/}

\newif\ifEnableSectionTOCLinks
\EnableSectionTOCLinksfalse 

\usepackage[top=1in, bottom=1.25in, left=1in, right=1in]{geometry}

\usepackage{microtype}
\usepackage{lineno}  
\usepackage{xspace} 
\usepackage{caption} 

\usepackage{graphicx}  
\usepackage{color}
\usepackage{colortbl}
\graphicspath{{./figs/}} 

\usepackage{amsmath} 
\usepackage{amssymb}
\usepackage{amsfonts}
\usepackage{upgreek} 

\newcommand*\patchAmsMathEnvironmentForLineno[1]{%
\expandafter\let\csname old#1\expandafter\endcsname\csname #1\endcsname
\expandafter\let\csname oldend#1\expandafter\endcsname\csname
end#1\endcsname
 \renewenvironment{#1}%
   {\linenomath\csname old#1\endcsname}%
   {\csname oldend#1\endcsname\endlinenomath}%
}
\newcommand*\patchBothAmsMathEnvironmentsForLineno[1]{%
  \patchAmsMathEnvironmentForLineno{#1}%
  \patchAmsMathEnvironmentForLineno{#1*}%
}
\AtBeginDocument{%
\patchBothAmsMathEnvironmentsForLineno{equation}%
\patchBothAmsMathEnvironmentsForLineno{align}%
\patchBothAmsMathEnvironmentsForLineno{flalign}%
\patchBothAmsMathEnvironmentsForLineno{alignat}%
\patchBothAmsMathEnvironmentsForLineno{gather}%
\patchBothAmsMathEnvironmentsForLineno{multline}%
\patchBothAmsMathEnvironmentsForLineno{eqnarray}%
}

\usepackage{hyperxmp}

\usepackage[pdftex,
            pdfauthor={\paperauthors},
            pdftitle={\paperasciititle},
            pdfkeywords={\paperkeywords},
            pdfcopyright={Copyright (C) \papercopyright},
            pdflicenseurl={\paperlicenceurl}]{hyperref}

\usepackage[colorinlistoftodos,textsize=scriptsize]{todonotes}

\usepackage[bottom,flushmargin,hang,multiple]{footmisc}

\usepackage[all]{hypcap} 

\usepackage{xspace} 
\usepackage{upgreek}

\def\lhcb   {\mbox{LHCb}\xspace}

\def\MagUp {\mbox{\em Mag\kern -0.05em Up}\xspace}

\ifthenelse{\boolean{uprightparticles}}%
{

 \def\Pmu         {\ensuremath{\upmu}\xspace}

 \def\Ppi         {\ensuremath{\uppi}\xspace}                 
                  
 \def\Prho        {\ensuremath{\uprho}\xspace}                 
                  
 \def\Ptau        {\ensuremath{\uptau}\xspace}

 \def\Ppsi        {\ensuremath{\uppsi}\xspace}

 \def\PDelta      {\ensuremath{\Delta}\xspace}                 
 \def\PXi         {\ensuremath{\Xi}\xspace}                 
 \def\PLambda     {\ensuremath{\Lambda}\xspace}                 
 \def\PSigma      {\ensuremath{\Sigma}\xspace}                 
 \def\POmega      {\ensuremath{\Omega}\xspace}                 
 \def\PUpsilon    {\ensuremath{\Upsilon}\xspace}
 \let\oldPi\Pi
 \def\PPi         {\ensuremath{\oldPi}\xspace}

 \def\PB      {\ensuremath{\mathrm{B}}\xspace}                 
 \def\PD      {\ensuremath{\mathrm{D}}\xspace}                 
 \def\PJ      {\ensuremath{\mathrm{J}}\xspace}                 
 \def\PK      {\ensuremath{\mathrm{K}}\xspace}                 
 \def\Pb      {\ensuremath{\mathrm{b}}\xspace}                 
 \def\Pc      {\ensuremath{\mathrm{c}}\xspace}                 
                  
 \def\Pe      {\ensuremath{\mathrm{e}}\xspace}                 
 \def\Ps      {\ensuremath{\mathrm{s}}\xspace}

 \def\thebaroffset{0.0em}
}
{

 \def\Pmu         {\ensuremath{\mu}\xspace}

 \def\Ppi         {\ensuremath{\pi}\xspace}                 
                  
 \def\Prho        {\ensuremath{\rho}\xspace}                 
                  
 \def\Ptau        {\ensuremath{\tau}\xspace}

 \def\Ppsi        {\ensuremath{\psi}\xspace}                 
                  
 \mathchardef\PDelta="7101
 \mathchardef\PXi="7104
 \mathchardef\PLambda="7103
 \mathchardef\PSigma="7106
 \mathchardef\POmega="710A
 \mathchardef\PUpsilon="7107
 \mathchardef\PPi="7105
 \def\PB      {\ensuremath{B}\xspace}                 
 \def\PD      {\ensuremath{D}\xspace}                 
 \def\PJ      {\ensuremath{J}\xspace}                 
 \def\PK      {\ensuremath{K}\xspace}                 
 \def\Pb      {\ensuremath{b}\xspace}                 
 \def\Pc      {\ensuremath{c}\xspace}                 
                  
 \def\Pe      {\ensuremath{e}\xspace}                 
 \def\Ps      {\ensuremath{s}\xspace}

 \def\thebaroffset{0.18em}
}
\newcommand{\offsetoverline}[2][\thebaroffset]{\kern #1\overline{\kern -#1 #2}}%

\makeatletter
\ifcase \@ptsize \relax
  \newcommand{\miniscule}{\@setfontsize\miniscule{4}{5}}
\or
  \newcommand{\miniscule}{\@setfontsize\miniscule{5}{6}}
\or
  \newcommand{\miniscule}{\@setfontsize\miniscule{5}{6}}
\fi
\makeatother

\DeclareRobustCommand{\optbar}[1]{\shortstack{{\miniscule (\rule[.5ex]{1.25em}{.18mm})}
  \\ [-.7ex] $#1$}}

\def\epem       {{\ensuremath{\Pe^+\Pe^-}}\xspace}

\def\mup        {{\ensuremath{\Pmu^+}}\xspace}

\def\mumu       {{\ensuremath{\Pmu^+\Pmu^-}}\xspace}

\def\tautau     {{\ensuremath{\Ptau^+\Ptau^-}}\xspace}

\def\squark    {{\ensuremath{\Ps}}\xspace}

\def\cquark    {{\ensuremath{\Pc}}\xspace}

\def\bquark    {{\ensuremath{\Pb}}\xspace}

\def\pion   {{\ensuremath{\Ppi}}\xspace}
\def\piz    {{\ensuremath{\pion^0}}\xspace}
\def\pip    {{\ensuremath{\pion^+}}\xspace}
\def\pim    {{\ensuremath{\pion^-}}\xspace}

\def\rhomeson {{\ensuremath{\Prho}}\xspace}

\def\rhop     {{\ensuremath{\rhomeson^+}}\xspace}

\def\kaon    {{\ensuremath{\PK}}\xspace}

\def\KorKbar {\kern \thebaroffset\optbar{\kern -\thebaroffset \PK}{}\xspace}

\def\Kp      {{\ensuremath{\kaon^+}}\xspace}

\def\D       {{\ensuremath{\PD}}\xspace}

\def\DorDbar {\kern \thebaroffset\optbar{\kern -\thebaroffset \PD}\xspace}

\def\Dp      {{\ensuremath{\D^+}}\xspace}
\def\Dm      {{\ensuremath{\D^-}}\xspace}

\def\DpDm    {\ensuremath{\Dp {\kern -0.16em \Dm}}\xspace}

\def\Dstarz  {{\ensuremath{\D^{*0}}}\xspace}

\def\B       {{\ensuremath{\PB}}\xspace}

\def\BorBbar {\kern \thebaroffset\optbar{\kern -\thebaroffset \PB}\xspace}
\def\Bz      {{\ensuremath{\B^0}}\xspace}

\def\Bd      {{\ensuremath{\B^0}}\xspace}

\def\BdorBdbar {\kern \thebaroffset\optbar{\kern -\thebaroffset \Bd}\xspace}
\def\Bu      {{\ensuremath{\B^+}}\xspace}

\def\Bs      {{\ensuremath{\B^0_\squark}}\xspace}

\def\BsorBsbar {\kern \thebaroffset\optbar{\kern -\thebaroffset \Bs}\xspace}
\def\Bc      {{\ensuremath{\B_\cquark^+}}\xspace}

\def\Bds     {{\ensuremath{\B_{(\squark)}^0}}\xspace}

\def\BdorBs  {\Bds}

\def\BdorBdstar  {{\ensuremath{\B^{(*)0}}}\xspace}
\def\Bsstar  {{\ensuremath{\B_{\squark}^{*0}}}\xspace}
\def\BsorBsstar  {{\ensuremath{\B_{\squark}^{(*)0}}}\xspace}
\def\Bdsstar {{\ensuremath{\B_{(\squark)}^{*0}}}\xspace}
\def\BdorBsstar {\Bdsstar}

\def\jpsi     {{\ensuremath{{\PJ\mskip -3mu/\mskip -2mu\Ppsi}}}\xspace}
\def\psitwos  {{\ensuremath{\Ppsi{(2S)}}}\xspace}

\def\Y#1S{\ensuremath{\PUpsilon{(#1S)}}\xspace}

\def\LorLbar     {\kern \thebaroffset\optbar{\kern -\thebaroffset \PLambda}\xspace}

\def\BF         {{\ensuremath{\mathcal{B}}}\xspace}

\newcommand{\decay}[2]{\ensuremath{#1\!\to #2}\xspace} 

\def\to                 {\ensuremath{\rightarrow}\xspace}

\def\AT#1     {\ensuremath{A_{\mathrm{T}}^{#1}}\xspace}           

\def\C#1      {\ensuremath{\mathcal{C}_{#1}}\xspace}                       
\def\Cp#1     {\ensuremath{\mathcal{C}_{#1}^{'}}\xspace}                    
\def\Ceff#1   {\ensuremath{\mathcal{C}_{#1}^{\mathrm{(eff)}}}\xspace}        
\def\Cpeff#1  {\ensuremath{\mathcal{C}_{#1}^{'\mathrm{(eff)}}}\xspace}       
\def\Ope#1    {\ensuremath{\mathcal{O}_{#1}}\xspace}                       
\def\Opep#1   {\ensuremath{\mathcal{O}_{#1}^{'}}\xspace}                    

\newcommand{\nospaceunit}[1]{\ensuremath{\text{#1}}}       
\newcommand{\aunit}[1]{\ensuremath{\text{\,#1}}}       

\newcommand{\tev}{\aunit{Te\kern -0.1em V}\xspace}
\newcommand{\gev}{\aunit{Ge\kern -0.1em V}\xspace}
\newcommand{\mev}{\aunit{Me\kern -0.1em V}\xspace}
\newcommand{\kev}{\aunit{ke\kern -0.1em V}\xspace}
\newcommand{\ev}{\aunit{e\kern -0.1em V}\xspace}
 
\newcommand{\mevc}{\ensuremath{\aunit{Me\kern -0.1em V\!/}c}\xspace}
\newcommand{\gevc}{\ensuremath{\aunit{Ge\kern -0.1em V\!/}c}\xspace}
\newcommand{\mevcc}{\ensuremath{\aunit{Me\kern -0.1em V\!/}c^2}\xspace}
\newcommand{\gevcc}{\ensuremath{\aunit{Ge\kern -0.1em V\!/}c^2}\xspace}

\def\mum  {\ensuremath{\,\upmu\nospaceunit{m}}\xspace}

\def\fb   {\ensuremath{\aunit{fb}}\xspace}
\def\invfb   {\ensuremath{\fb^{-1}}\xspace}

\def\gsim{{~\raise.15em\hbox{$>$}\kern-.85em
          \lower.35em\hbox{$\sim$}~}\xspace}
\def\lsim{{~\raise.15em\hbox{$<$}\kern-.85em
          \lower.35em\hbox{$\sim$}~}\xspace}

\def\pt         {\ensuremath{p_{\mathrm{T}}}\xspace}

\def\ptot       {\ensuremath{p}\xspace}

\def\evtgen     {\mbox{\textsc{EvtGen}}\xspace}

\def\geant      {\mbox{\textsc{Geant4}}\xspace}

\def\photos     {\mbox{\textsc{Photos}}\xspace}

\def\pythia     {\mbox{\textsc{Pythia}}\xspace}

\def\tell1  {TELL1\xspace}
\def\ukl1   {UKL1\xspace}

\newcommand{\lhcborcid}[1]{\href{https://orcid.org/#1}{\hspace*{0.1em}\raisebox{-0.45ex}{\includegraphics[width=1em]{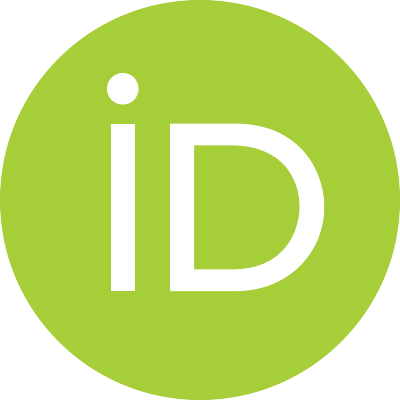}}}}

\hypersetup{
  colorlinks   = true, 
  urlcolor     = blue, 
  linkcolor    = blue, 
  citecolor    = red   
}

\ifEnableSectionTOCLinks
    \usepackage[explicit]{titlesec} 
    
    \let\oldcontentsline\contentsline
    \renewcommand

    \titleformat{\section}{\normalfont\Large\bf}{\hyperlink{tocsection.\thesection}{{\thesection} \parbox[t]{\dimexpr\textwidth-1pc}{#1}}}{1pc}{}

    \titleformat{\subsection}{\normalfont\bf}{\hyperlink{tocsubsection.\thesubsection}{{\thesubsection} \parbox[t]{\dimexpr\textwidth-1pc}{#1}}}{1pc}{}

    \titleformat{name=\section,numberless}[display]{}{}{0pt}{\normalfont\Huge\bfseries #1}
\fi

\usepackage{cite} 
\usepackage{mciteplus}

\usepackage{longtable} 

\def\Bstarz      {{\ensuremath{\B^{*0}}}\xspace}
\def\Bsstar  {{\ensuremath{\B^{*0}_\squark}}\xspace}
\def\BorBs  {{\ensuremath{\B^{0}_{(\squark)}}}\xspace}
\def\BorBsstar  {{\ensuremath{\B^{*0}_{(\squark)}}}\xspace}
\begin{document}

\renewcommand{\thefootnote}{\fnsymbol{footnote}}
\setcounter{footnote}{1}


\begin{titlepage}
\pagenumbering{roman}

\vspace*{-1.5cm}
\centerline{\large EUROPEAN ORGANIZATION FOR NUCLEAR RESEARCH (CERN)}
\vspace*{1.5cm}
\noindent
\begin{tabular*}{\linewidth}{lc@{\extracolsep{\fill}}r@{\extracolsep{0pt}}}
\ifthenelse{\boolean{pdflatex}}
{\vspace*{-1.5cm}\mbox{\!\!\!\includegraphics[width=.14\textwidth]{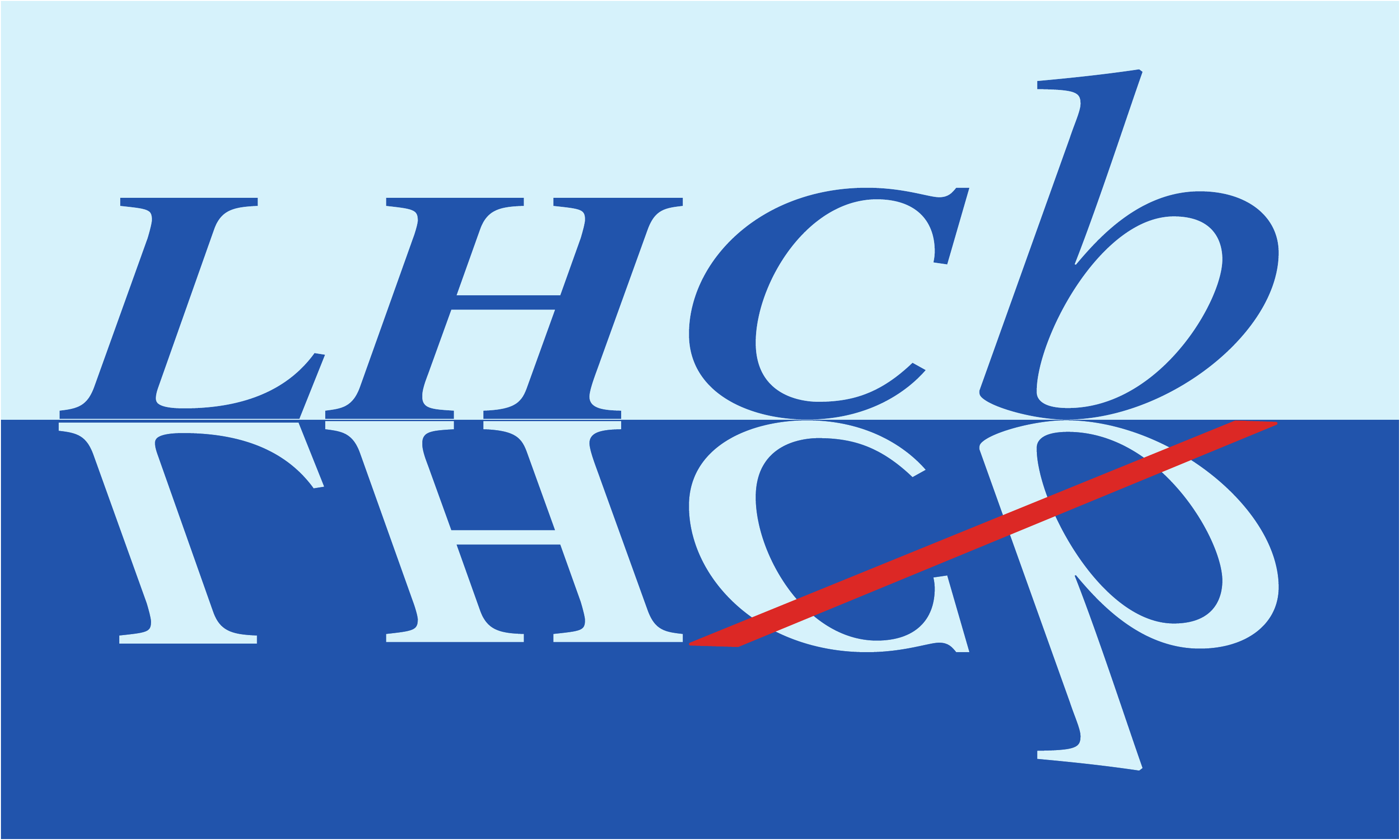}} & &}%
{\vspace*{-1.2cm}\mbox{\!\!\!\includegraphics[width=.12\textwidth]{figs/lhcb-logo.eps}} & &}%
\\
 & & CERN-EP-2024-220 \\  
 & & LHCb-PAPER-2024-026 \\  
 & & January 14, 2025\\
 & & \\
\end{tabular*}

\vspace*{2.0cm}

{\normalfont\bfseries\boldmath\huge
\begin{center}
  \papertitle 
\end{center}
}

\vspace*{1.0cm}

\begin{center}
\paperauthors\footnote{Authors are listed at the end of this paper.}
\end{center}

\vspace{\fill}

\begin{abstract}
  \noindent
A search for the very rare $B^{*0}\to\mu^+\mu^-$ and $B_{s}^{*0}\to\mu^+\mu^-$ decays is conducted by analysing the $B_c^+\to \pi^+\mu^+\mu^-$ process. 
The analysis uses proton-proton collision data collected with the LHCb detector between 2011 and 2018, corresponding to an integrated luminosity of 9$\text{\,fb}^{-1}$. 
The signal signatures correspond to simultaneous peaks in the $\mu^+\mu^-$ and $\pi^+\mu^+\mu^-$ invariant masses. 
No evidence for an excess of events over background is observed for either signal decay mode. 
Upper limits at the $90\%$ confidence level are set on the branching fractions relative to that for $B_c^+\to J\mskip -3mu/\mskip -2mu\psi\pi^+$ decays,  
\begin{align*}
    {\cal R}_{B^{*0}(\mu^+\mu^-)\pi^+/J\mskip -3mu/\mskip -2mu\psi\pi^+} &<  3.8\times 10^{-5}\ \text{ and }\\ 
    {\cal R}_{B_{s}^{*0}(\mu^+\mu^-)\pi^+/J\mskip -3mu/\mskip -2mu\psi\pi^+} &< 5.0\times 10^{-5}\,. 
\end{align*}
  
\end{abstract}

\vspace*{1.0cm}

\begin{center}
Published in \href{https://doi.org/10.1140/epjc/s10052-024-13573-0}{Eur. Phys. J. C 85, 20 (2025)}
\end{center}

\vspace{\fill}

{\footnotesize 
\centerline{\copyright~\papercopyright. \href{\paperlicenceurl}{\paperlicence}.}}
\vspace*{2mm}

\end{titlepage}


\newpage
\setcounter{page}{2}
\mbox{~}
%
%
%
%


\renewcommand{\thefootnote}{\arabic{footnote}}
\setcounter{footnote}{0}

\cleardoublepage


\pagestyle{plain} 
\setcounter{page}{1}
\pagenumbering{arabic}


\section{Introduction}
\label{sec:Introduction}

Weak decays of the \Bstarz and \Bsstar excited vector mesons into leptonic final states offer the opportunity to search for possible deviations from Standard Model~(SM) expectations. 
Unlike the weak leptonic decays of the \Bz and \Bs pseudoscalar mesons, the decays of excited vector mesons are not suppressed by the chiral structure of the SM weak interaction~\cite{Buchalla:1995vs,Grinstein:2015aua,Khodjamirian:2015dda}. 
However, since the \BdorBsstar mesons decay predominantly through the electromagnetic interaction, the branching fractions for their weak leptonic decays are highly suppressed in the SM.
For example, the \decay{\Bsstar}{\mumu} branching fraction is expected in the SM to be around $10^{-11}$~\cite{Grinstein:2015aua,Khodjamirian:2015dda}, but could be enhanced due to physics beyond the SM. 
The impact of particular extensions of the SM on the leptonic decays of \Bstarz and \Bsstar mesons has been investigated in Refs.~\cite{Xu:2015eev,Sahoo:2016edx,Banerjee:2017upm,Kumar:2017xgl,Kumbhakar:2018uty}.

Many experimental studies of the \BdorBs\ leptonic decays have been performed. 
Recent results include measurements of the \decay{\Bs}{\mumu} branching fraction and limits on the \decay{\Bz}{\mumu} rate that are consistent with SM expectations~\cite{LHCb-PAPER-2021-007,LHCb-PAPER-2021-008,CMS:2022mgd,ATLAS:2018cur}, as well as limits on the rates of \decay{\BdorBs}{\epem} and \decay{\BdorBs}{\tautau} decays~\cite{LHCb-PAPER-2020-001,LHCb-PAPER-2017-003}.
However, there has not yet been any search for a \decay{\BdorBsstar}{\ell^+\ell^-} decay mode. 
In this paper, the first search for the \decay{\Bstarz}{\mumu} and \decay{\Bsstar}{\mumu} decays is presented.
The analysis is based on the data samples collected with the LHCb detector between 2011 and 2018, corresponding to an integrated luminosity of $9\invfb$ of proton-proton ($pp$) collisions at centre-of-mass energies of $7$, $8$ and $13 \tev$.
As discussed in Ref.~\cite{Abudinen:2022dnw}, searches via prompt \BorBsstar~production in LHC collisions are expected to be limited by the large amount of background from $pp$ interactions. 
The search is therefore performed via the \decay{\Bc}{\BorBsstar\pip}, \decay{\BorBsstar}{\mumu} decay chain, subsequently denoted as \decay{\Bc}{\BorBsstar(\mumu)\pip} decays.
This is expected to be the most promising method as it exploits the displaced \Bc-vertex signature to suppress background; a similar approach has recently been demonstrated in a search for the \decay{\Dstarz}{\mumu} decay~\cite{LHCb-PAPER-2023-004}. 
The inclusion of charge-conjugate processes is implied throughout the paper. 

The analysis follows procedures from a recent search for nonresonant \mbox{\decay{\Bc}{\pip\mumu}} decays~\cite{LHCb-PAPER-2023-037}.
The results of that analysis include an upper limit on the ratio \mbox{$\BF(\decay{\Bc}{\pip\mumu})/\BF(\decay{\Bc}{\jpsi\pip})<1.9\times 10^{-4}$} at $90\%$ confidence level~(CL) in the interval  $15.0 < q^2 < 35.0\gev^2/c^4$, where $q^2 = m^2(\mumu)$ is the square of the invariant mass of the dimuon system. 
That result can be used to set limits on the branching fraction products ${\cal B}\left(\decay{\Bc}{\Bsstar\pip}\right) \times {\cal B}\left(\decay{\Bsstar}{\mumu}\right)$ and ${\cal B}\left(\decay{\Bc}{\Bstarz\pip}\right) \times {\cal B}\left(\decay{\Bstarz}{\mumu}\right)$, since such decays would contribute in the relevant $q^2$ region.
However, due to the narrow $\BdorBsstar$ width, significantly better experimental sensitivity can be obtained by a dedicated search with optimised selection requirements and fit strategy, as presented here.
The previous result also implies that the contribution from nonresonant \decay{\Bc}{\pip\mumu}~decays is negligible, and therefore these decays do not need to be considered as a source of background in the data used for the \BorBsstar search.

To search for \decay{\Bc}{\BorBsstar(\mumu)\pip} signals, the reconstructed $\Bc$-candidate invariant mass, $m(\mumu\pip)$, and the dimuon invariant mass, $m(\mumu)$, serve as discriminating observables in an extended unbinned maximum-likelihood fit. 
The analysis uses the \mbox{\decay{\Bc}{\jpsi(\mumu)\,\pip}}~decay as a normalisation mode. 
The signal yields, relative to that for the normalisation mode, are translated into branching fraction ratios through
\begin{align}
    {\cal R}_{\BorBsstar(\mumu)\pip/\jpsi\pip} &\equiv \frac{{\cal B}(\decay{\Bc}{\BorBsstar(\mumu)\pip})}{{\cal B}(\decay{\Bc}{\jpsi\pip})}\, \nonumber \\
    &= \frac{N_{\BorBsstar\pip}}{N_{\jpsi\pip}} \cdot \frac{\varepsilon_{\jpsi\pip}}{\varepsilon_{\BorBsstar\pip}} \cdot {\cal B}(\decay{\jpsi}{\mumu})\, \nonumber \\
    &= \alpha_{\BorBsstar\pip}^{\rm SES}\cdot N_{\BorBsstar\pip}\,,
   \label{eq:R_BF_norm} 
\end{align}
where $N$ denotes the yield of the mode indicated in the subscript, $\varepsilon$ denotes the corresponding efficiency determined from simulation with data-driven corrections, and ${\cal B}(\decay{\jpsi}{\mumu})$ is the known branching fraction of the \decay{\jpsi}{\mumu} decay~\cite{PDG2024}.
The single-event sensitivity $\alpha_{\BorBsstar\pip}^{\rm SES}$ is the value of the ratio that would be obtained for one signal decay.

\section{Detector and simulation} 
\label{sec:Detector} 

The \lhcb detector~\cite{LHCb-DP-2008-001,LHCb-DP-2014-002} is a single-arm forward
spectrometer covering the \mbox{pseudorapidity} range $2<\eta <5$,
designed for the study of particles containing \bquark or \cquark~quarks. 
The detector includes a high-precision tracking system
consisting of a silicon-strip vertex detector surrounding the $pp$
interaction region~\cite{LHCb-DP-2014-001}, a large-area silicon-strip detector located
upstream of a dipole magnet with a bending power of about
$4{\mathrm{\,T\,m}}$, and three stations of silicon-strip detectors and straw
drift tubes~\cite{LHCb-DP-2013-003,LHCb-DP-2017-001}
placed downstream of the magnet.
The tracking system provides a measurement of the momentum, \ptot, of charged particles with
a relative uncertainty that varies from 0.5\% at low momentum to 1.0\% at 200\gevc.
The minimum distance of a track to a primary $pp$ collision vertex~(PV), the impact parameter (IP), 
is measured with a resolution of $(15+29/\pt)\mum$,
where \pt is the component of the momentum transverse to the beam, in\,\gevc.
Different types of charged hadrons are distinguished using information
from two ring-imaging Cherenkov detectors~\cite{LHCb-DP-2012-003}. 
Photons, electrons and hadrons are identified by a calorimeter system consisting of
scintillating-pad and preshower detectors, an electromagnetic
and a hadronic calorimeter. Muons are identified by a
system composed of alternating layers of iron and multiwire
proportional chambers~\cite{LHCb-DP-2012-002}.

The online event selection is performed by a trigger~\cite{LHCb-DP-2012-004,LHCb-DP-2019-001}, 
which consists of a hardware stage, based on information from the calorimeter and muon
systems, followed by a two-level software stage, which reconstructs the full event.
Candidate $\Bc\to\pip\mumu$ decays are triggered as described in Ref.~\cite{LHCb-PAPER-2023-004} for \Bu\ decays to the same final states. 
The hardware stage of the trigger selects events containing at least one muon with high $\pt$. 
The following software stage selects events containing at least one high-\pt muon detached from any~PV. 
The events must contain at least one secondary vertex (formed by two or more of the final-state particles) that is also detached from any PV. 
Secondary vertices consistent with the decay of a \bquark~hadron are identified by multivariate algorithms~\cite{BBDT,LHCb-PROC-2015-018}. 

Simulation is used to optimise the event selection procedure, to model the shape of invariant-mass distributions and to estimate efficiencies accounting for the effects of the detector acceptance, reconstruction and selection criteria. 
In the simulation, $pp$ collisions are generated using \pythia~\cite{Sjostrand:2007gs} with a specific \lhcb configuration~\cite{LHCb-PROC-2010-056}. 
The production of \Bc~mesons is simulated using the dedicated generator {\textsc BcVegPy}~\cite{Chang:2005hq}.
Decays of unstable particles are described by \evtgen~\cite{Lange:2001uf}, in which final-state radiation is generated using \photos~\cite{davidson2015photos}.
The interaction of the generated particles with the detector, and its response, are implemented using the \geant toolkit~\cite{Allison:2006ve, *Agostinelli:2002hh, LHCb-PROC-2011-006, LHCb-DP-2018-004}. 

The \Bc candidates reconstructed in simulation are weighted to correct for discrepancies between data and simulation associated with the particle-identification~\cite{LHCb-DP-2018-001}, track-reconstruction~\cite{LHCb-DP-2013-002} and hardware-trigger~\cite{LHCb-DP-2012-004} efficiencies.
The simulation is also corrected such that the \Bc~lifetime corresponds to the current experimental value~\cite{PDG2024,LHCb-PAPER-2013-063,LHCb-PAPER-2014-060}. 
Additional corrections are applied to account for discrepancies in \Bc production kinematics, event track multiplicity and other observables used in the selection of \Bc~candidates.
These corrections are obtained using a multivariate weighting algorithm~\cite{Rogozhnikov:2016bdp}, which is trained using \mbox{\decay{\Bc}{\jpsi\pip}} decays in background-subtracted data and simulation. 
After the corrections are applied, the simulated distributions of all variables used in the analysis are in good agreement with the data.

\section{Candidate selection and background sources} 
\label{sec:Selection} 

The initial stages of the offline selection are identical to those for the recent search for nonresonant \decay{\Bc}{\pip\mumu} decays~\cite{LHCb-PAPER-2023-037}.
The \Bc~candidates are formed from pairs of oppositely charged tracks identified as muons together with a track identified as a pion.
The tracks are required to form a vertex with a good kinematic-fit quality that is displaced from every PV.
Each \Bc~candidate must have a momentum vector that is aligned with the direction between one of the PVs, subsequently referred to as the associated PV, and the \Bc-candidate decay vertex.

Each \Bc candidate is required to have an invariant mass in the range \mbox{$6150 < m(\pip\mumu) < 6700 \mevcc$}. 
The expected signal resolution in $m(\pip\mumu)$ is about $20\mevcc$.
The dimuon invariant mass is calculated from the outcome of a kinematic fit in which the \Bc-candidate invariant mass is constrained to the known \Bc~mass~\cite{PDG2024} and the momentum vector is constrained to be consistent with the line of flight between the associated PV and the decay vertex, thereby improving the resolution. 
The dimuon invariant mass is required to be in the range $5225 < m(\mumu) < 5515 \mevcc$ for the signal modes and $3000 < m(\mumu) < 3200 \mevcc$ for the normalisation mode. 
The expected signal resolution in $m(\mumu)$ is about $4\mevcc$.

Combinatorial background arising from random combinations of tracks is suppressed using a boosted decision tree~(BDT) classifier~\cite{Breiman,Chen:2016:XST:2939672.2939785} that has been trained and validated to identify \decay{\Bc}{\pip\mumu} signal candidates, such that its performance is independent of the dimuon invariant mass~\cite{LHCb-PAPER-2023-037}.
The BDT classifier receives as inputs the \pt of the pion track, the highest \pt among muon tracks, the IPs of the muon tracks and of the \Bc~candidate with respect to the associated PV, the \Bc flight distance, the vertex quality of the \Bc~candidate, and the largest distance of closest approach between any two of the final-state particles.

Further suppression of combinatorial background is obtained by applying a requirement on the cosine of the helicity angle $\theta_l$, which is defined as the angle between the \mup direction and the direction opposite of the \Bc~momentum in the dimuon rest frame.
This has additional discrimination power since the signal and normalisation modes follow a $1-\cos^2{\theta_l}$ distribution while the combinatorial background sharply peaks at $\cos \theta_l \approx \pm 1$.  

Requirements on the BDT classifier output, the absolute value of $\cos{\theta_l}$, and variables characterising the charged-pion particle identification are optimised simultaneously.
The optimisation is based on a grid search to obtain the best signal sensitivity using the figure of merit $\varepsilon / (5/2 + \sqrt{N_{\rm B}})$~\cite{Punzi:2003bu}, 
where $\varepsilon$ is the signal efficiency and $N_{\rm B}$ is the expected number of background candidates in the signal region. 
The figure of merit is evaluated separately for \decay{\Bc}{\Bstarz(\mumu)\pip} and \decay{\Bc}{\Bsstar(\mumu)\pip} decays. 
The signal region for each decay mode corresponds to a two-dimensional range in $m(\pip\mumu)$ and $m(\mumu)$ of about \mbox{$\pm3$ times} the expected resolution in each dimension, centred at the expected two-dimensional peak position~\cite{PDG2024}.  
The expected background yield is estimated by fitting a background-only model to the dataset excluding the region \mbox{$ 6215 < m(\pip\mumu) < 6335\mevcc$}. 

The figures of merit for both \decay{\Bc}{\Bstarz(\mumu)\pip} and \decay{\Bc}{\Bsstar(\mumu)\pip} decays have maximum values at the same grid point. 
With the optimised requirements, the classifier has a combinatorial background rejection power of $99\%$, whilst retaining $65\%$ of signal decays.
The optimised angular selection, corresponding to $\left| \cos{\theta_l}\right| < 0.90$, further rejects about $30\%$ of the combinatorial background, whilst keeping about $98\%$ of signal decays. 
The particle-identification requirements have a pion efficiency around $90\%$, with a kaon misidentification rate around $10\%$.  
The particle-identification requirements applied to the muon candidates have an efficiency around $99\%$, with a pion misidentification rate below $1\%$. 
The same selection requirements are used for signal and normalisation modes to reduce potential systematic biases on the measurement of branching fraction ratios. 
After applying the selection requirements, each selected event contains only one \Bc~candidate. 

Backgrounds from partially reconstructed decays, such as \decay{\Bc}{\jpsi\rhop(\pip\piz)}~\cite{LHCb-PAPER-2023-046} for the normalisation mode, have a reconstructed \Bc-candidate invariant mass that lies more than $100\mevcc$ below the known \Bc~mass~\cite{PDG2024}. 
These sources of background predominantly populate a region outside, but have a tail that extends into, the fit range used in the analysis. 
This is also the case for backgrounds such as \mbox{\decay{\Bc}{\rhop\mumu}} for the signal modes, but these were found to be negligible in the search for nonresonant \decay{\Bc}{\pip\mumu} decays~\cite{LHCb-PAPER-2023-037}. 
The partially reconstructed background contribution is therefore neglected in the fit for the signal modes, but is accounted for in the normalisation mode fit.
Processes with a missing neutrino or two or more missing massive particles can also be a source of partially reconstructed background, but their contributions are negligible in the fit range. 

Contributions from hadronic backgrounds such as \mbox{\decay{\Bc}{\pip\pim\pip}} decays, where two pions are mistakenly identified as muons, were found to be negligible in the search for nonresonant \decay{\Bc}{\pip\mumu} decays~\cite{LHCb-PAPER-2023-037} and are therefore neglected.  
Similarly, possible contributions from the resonant \decay{\Bc}{\jpsi\pip} or \decay{\Bc}{\psitwos\pip} decays, where the pion is mistakenly identified as one of the two muons and vice versa, were studied using simulation and data and found to be negligible after applying the selection requirements. 
Contributions from \decay{\Bc}{\BorBs\pip \to \mumu \pip} decays are negligible due to their small rates~\cite{LHCb-PAPER-2021-007,LHCb-PAPER-2021-008,CMS:2022mgd,ATLAS:2018cur} and due to the selection requirements, which suppress topologies with a \mumu~vertex displaced from the \Bc-decay vertex. 

For the normalisation mode, misidentified background can arise from the \decay{\Bc}{\jpsi\Kp} mode.
The branching fraction for this decay is Cabibbo-suppressed with respect to that for the \decay{\Bc}{\jpsi\pip} decay, and their ratio has been measured to be $0.079\pm 0.007 \pm 0.003$~\cite{LHCb-PAPER-2016-020}.
This background is further suppressed by the particle-identification requirements, but nonetheless is accounted for in the normalisation mode fit.

\section{Invariant-mass fits} 
\label{sec:Fit} 

The yield of the \decay{\Bc}{\jpsi(\mumu)\,\pip} normalisation mode is determined from a one-dimensional extended unbinned maximum-likelihood fit to the $m(\pip\mumu)$ distribution of candidates in the range $3000 < m(\mumu) < 3200 \mevcc$. 
The normalisation mode also provides correction factors that account for discrepancies between data and simulation in the signal peak positions and widths. 
The relevant factors for the dimuon signal shape are obtained from an additional maximum-likelihood fit to the $m(\mumu)$ distribution. 
The \Bc-candidate invariant-mass and dimuon invariant-mass fits to the normalisation mode are independent of each other. 
A two-dimensional fit is avoided since possible correlations in the tail regions of the two observables could result in a non-negligible fit bias given the large sample size. 

For the \Bc-candidate invariant-mass fit for the normalisation mode, the fit model includes four components: signal \decay{\Bc}{\jpsi \pip} decays, misidentified \decay{\Bc}{\jpsi \Kp} decays, partially reconstructed background from \decay{\Bc}{\jpsi \rhop} decays and combinatorial background. 
The signal, misidentified and partially reconstructed backgrounds are each modelled by the sum of two Gaussian functions, one of which has power-law tails~\cite{Skwarnicki:1986xj}. 
The relative fraction between the two Gaussian functions and the tail parameters of each distribution are fixed from simulation. 
The peak position and width of the distributions are allowed to vary in the fit to the data by a global shift and scaling factor, respectively, that are shared between these three components. 
The combinatorial background model is an exponential function with a slope that is allowed to vary.
In total, the fit includes seven parameters: the yields of the four components, the global peak position shift and width scaling factor, and the slope of the combinatorial background. 
The yield for misidentified $\decay{\Bc}{\jpsi\Kp}$ decays is allowed to vary with respect to the yield for the $\decay{\Bc}{\jpsi\pip}$ decays within a Gaussian constraint based on the expected misidentification rate~\cite{LHCb-DP-2018-001} and the measured branching fraction ratio~\cite{LHCb-PAPER-2016-020}.

For the dimuon invariant-mass fit, the fit model includes  a signal and a combinatorial background component. 
The signal is modelled by a Gaussian function with power-law tails, while the background is modelled by a first-order polynomial function.
The tail parameters of the signal model are fixed from \decay{\Bc}{\jpsi\pip} simulation. The signal peak position and width are allowed to vary in the fit to the data through a shift and a width scaling factor.
The fit to the dimuon-mass includes five free parameters: the yields for the two components, the shift of peak position and width scaling factor, and the slope of the combinatorial background. 

\begin{figure}[tb]
\centering
\includegraphics[width=0.50\textwidth]{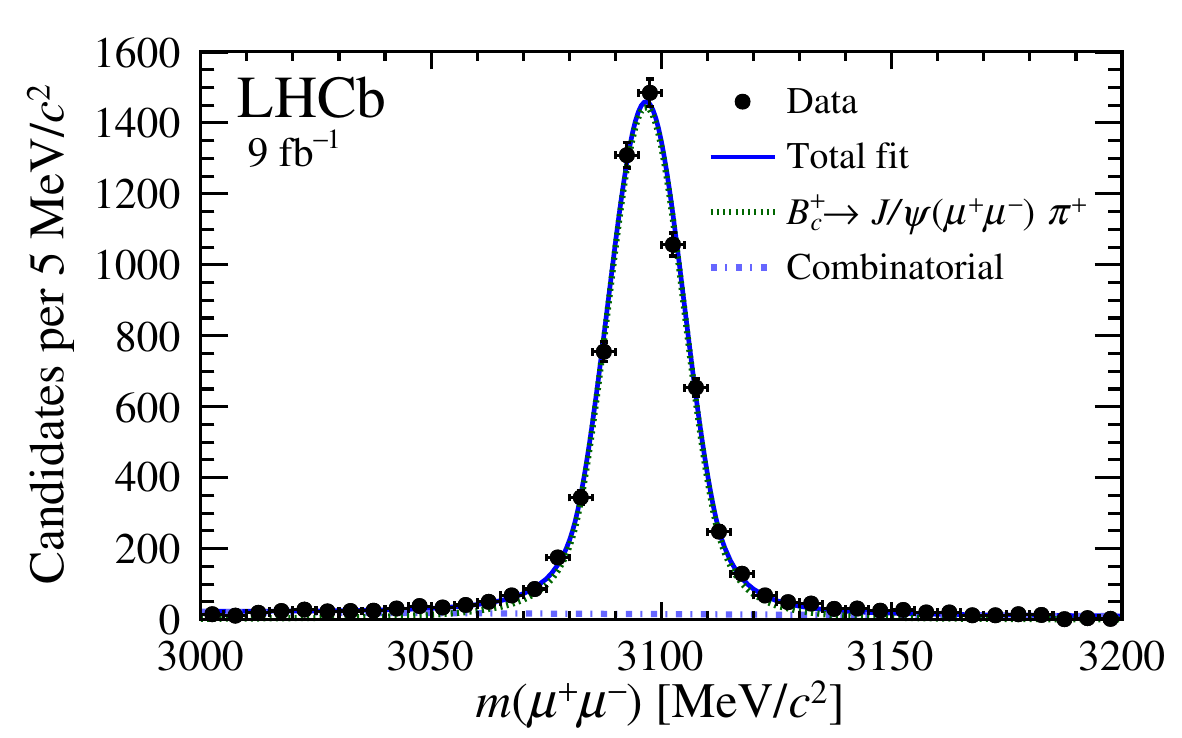}\hfill
\includegraphics[width=0.50\textwidth]{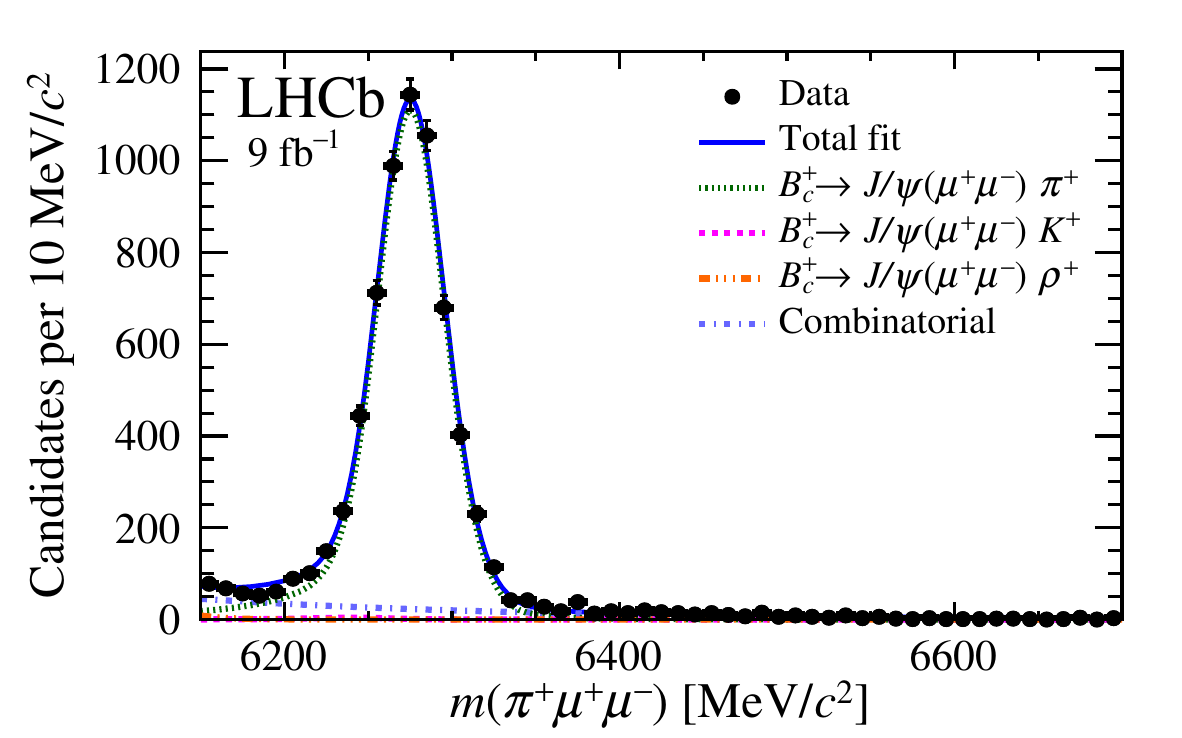}
\caption{
    Reconstructed (left) $\mu^+\mu^-$ and (right) $\pi^+\mu^+\mu^-$ invariant-mass distributions for the selected \mbox{$B_c^+\to  J\mskip -3mu/\mskip -2mu\psi(\mu^+\mu^-)\pi^+$} candidates, with results of the fit superimposed.
    The distributions are dominated by signal, and the background contributions are barely visible.
}
\label{fig:mass:norm} 
\end{figure}

Figure~\ref{fig:mass:norm} shows the dimuon and \Bc-candidate invariant-mass distributions of selected \mbox{\decay{\Bc}{\jpsi\pip}} candidates. 
The \Bc-candidate invariant-mass fit converges to a yield of $6213 \pm89$ decays, where the uncertainty is statistical only.

The dimuon invariant-mass distribution in data can receive contributions from \jpsi~decays that do not stem from the \mbox{\decay{\Bc}{\jpsi\pip}} process.
This background contribution could affect the fit results for the shift and width scaling factors.
To check this and also the effect of dependencies in the tails between \Bc-mass and dimuon mass distributions, the fit is repeated restricting the \Bc-candidates to the region \mbox{$6215 < m(\pip\mumu) < 6335\mevcc$}. 
The results for the shift and width scaling factor obtained from this fit demonstrate that the values obtained from the nominal fit are robust against systematic variations.

The signal \decay{\Bc}{\Bstarz(\mumu)\pip} and \decay{\Bc}{\Bsstar(\mumu)\pip} yields are determined from a two-dimensional extended unbinned maximum-likelihood fit to the $m(\mumu)$ and $m(\pip\mumu)$ distributions. 
The fit model includes three components: signal \mbox{\decay{\Bc}{\Bstarz(\mumu)\pip}} decays, signal \mbox{\decay{\Bc}{\Bsstar(\mumu)\pip}} decays and combinatorial background. 
For each component, the total model is the product of the respective dimuon and \Bc-candidate invariant-mass models.
The models for the signal components are validated using simulation. 
The two fit observables are found to not be significantly correlated in simulation or sideband data and are therefore treated as uncorrelated.  

For the signal components, the dimuon and the \Bc-candidate invariant-mass distributions are each modelled using a Gaussian function with power-law tails on both sides of the peak. 
The tail parameters are fixed to the values obtained from simulation. 
The signal dimuon and \Bc-candidate invariant-mass models each include a global shift of peak position and a global scaling factor for the width of the distribution, relative to the values found in simulation.
For the combinatorial background, the dimuon and the \Bc-candidate invariant-mass distributions are modelled using a linear function and an exponential function, respectively, 
with slope parameters allowed to vary in the fit to data. 

In total, the fit includes five free parameters: the yields for each component and the two parameters of the combinatorial background model.
In addition, the global peak position shift and width scaling factor for each of the dimuon and \Bc-candidate invariant-mass models are allowed to vary within Gaussian constraints based on the values obtained from the fits for the \decay{\Bc}{\jpsi(\mumu)\,\pip} candidates. 

\begin{figure}[tb]
\centering
\includegraphics[width=0.50\textwidth]{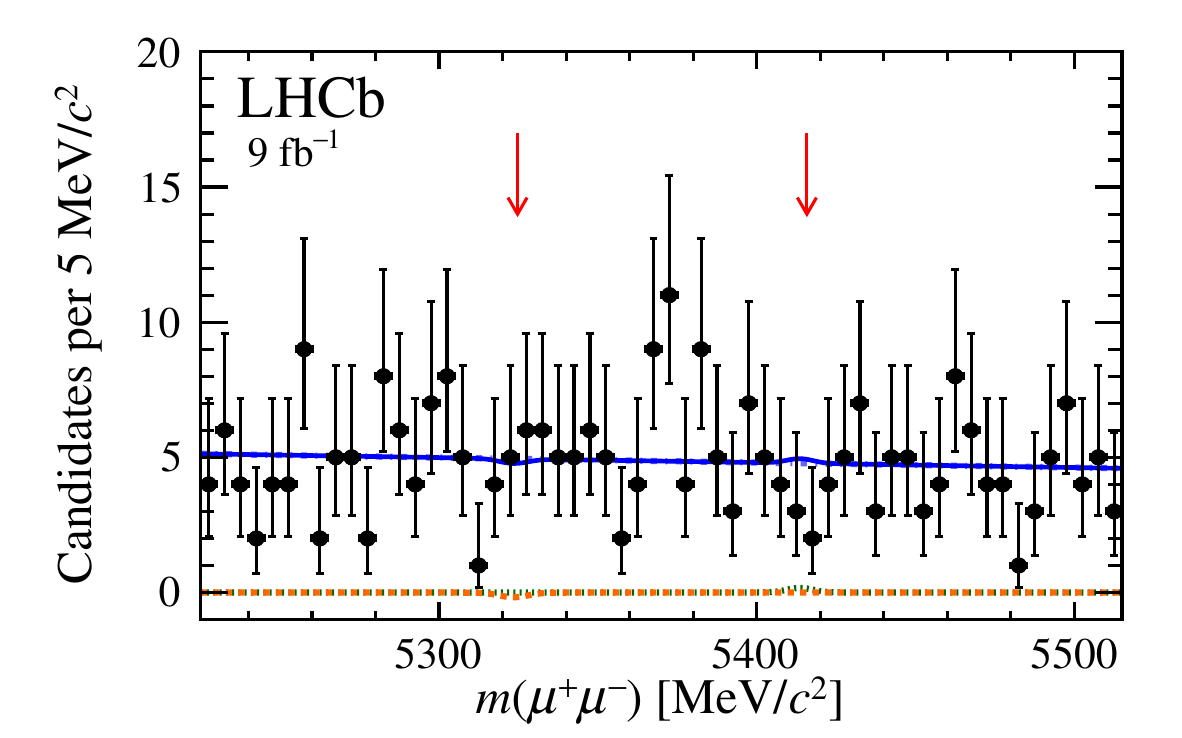}\hfill
\includegraphics[width=0.50\textwidth]{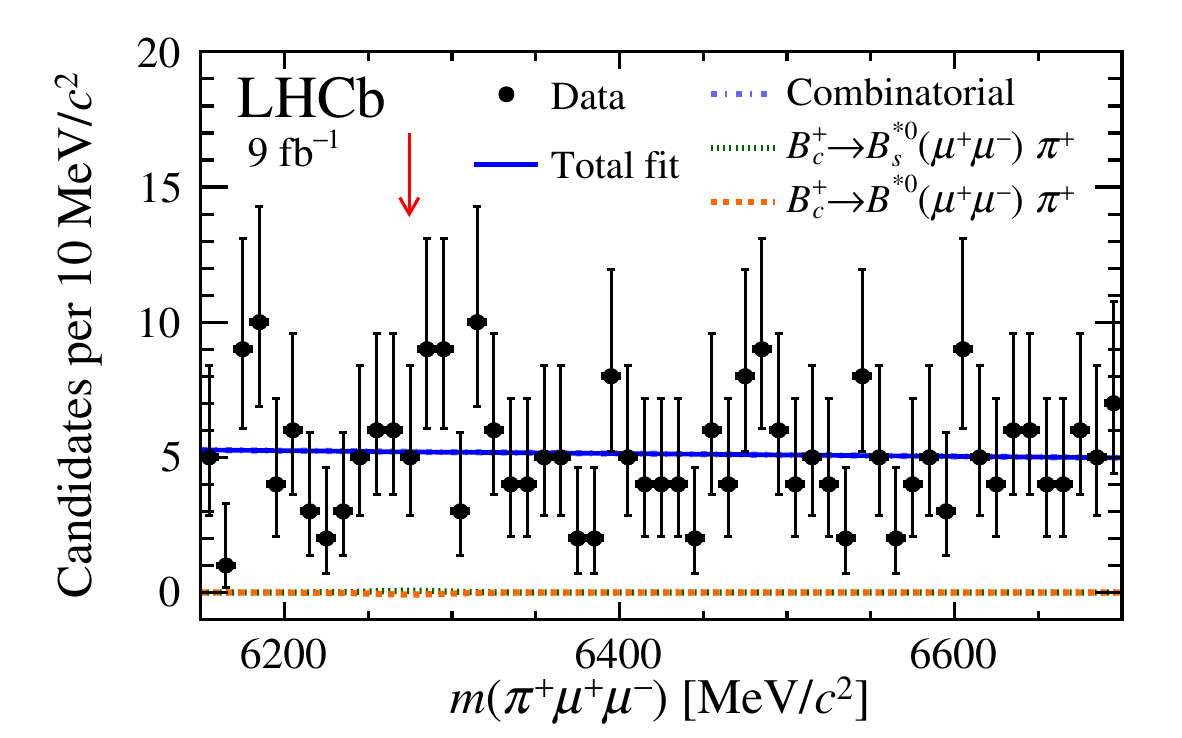} 
\caption{
    Reconstructed (left) $\mu^+\mu^-$ and (right) $\pi^+\mu^+\mu^-$ invariant-mass distributions for the selected \mbox{$B_c^+\to B_{(s)}^{*0}(\mu^+\mu^-)\pi^+$} candidates, with results of the fit superimposed.
    The red arrows point to the expected signal peak positions.
    The signal yields are consistent with zero and therefore the $B_c^+\to B^{*0}(\mu^+\mu^-)\pi^+$ and $B_c^+\to B_{s}^{*0}(\mu^+\mu^-)\pi^+$ components are barely visible. 
}
\label{fig:mass:signal} 
\end{figure}

\begin{figure}[tb]
    \centering
    \includegraphics[width=0.50\textwidth]{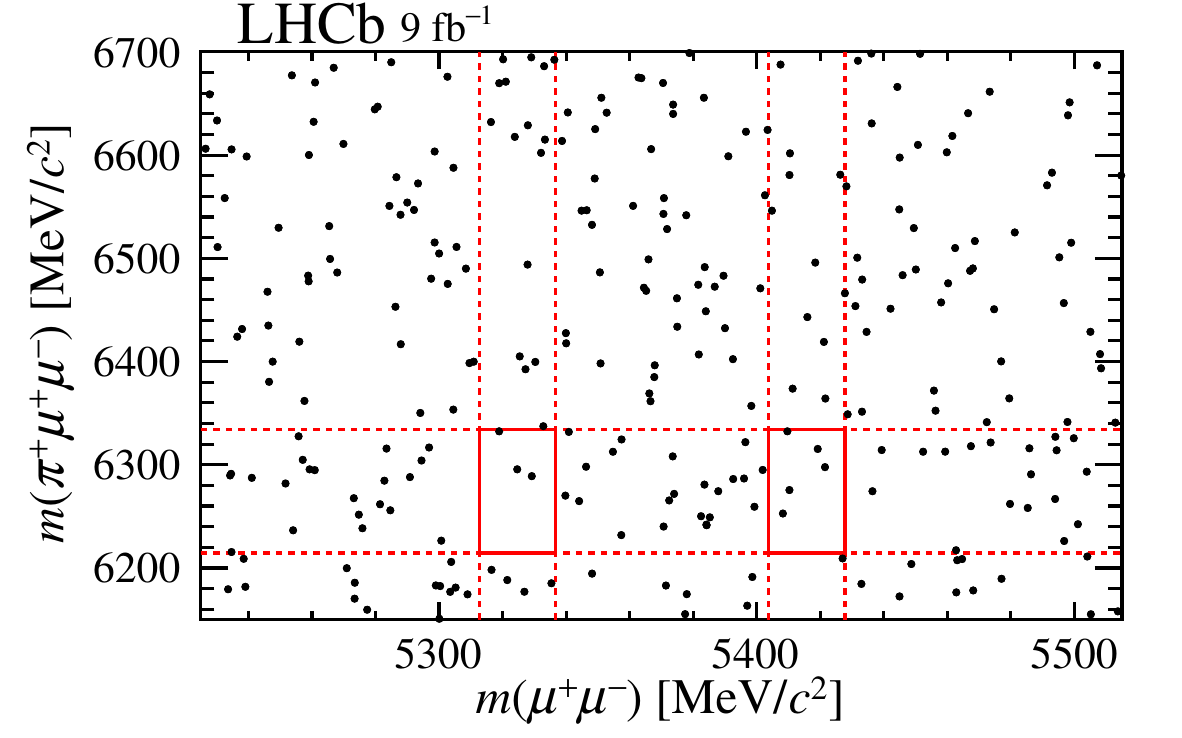}\hfill
    \includegraphics[width=0.50\textwidth]{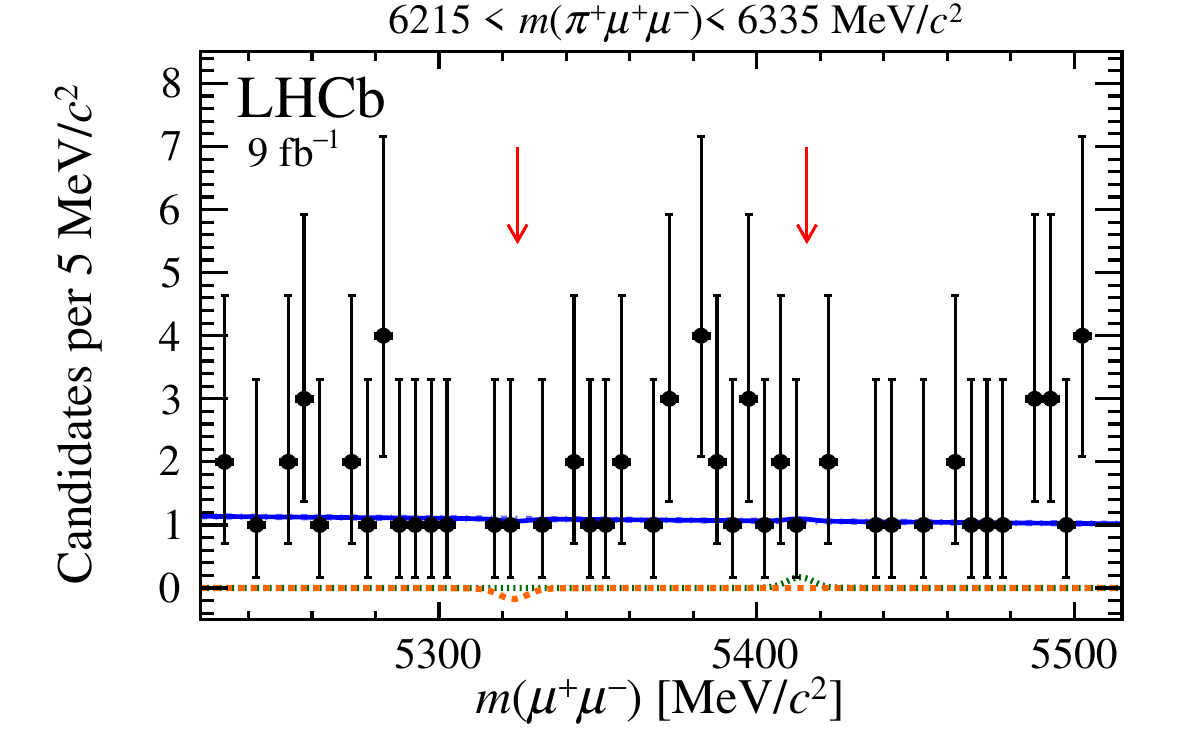}\\\vspace{0.1cm}
    
    \includegraphics[width=0.50\textwidth]{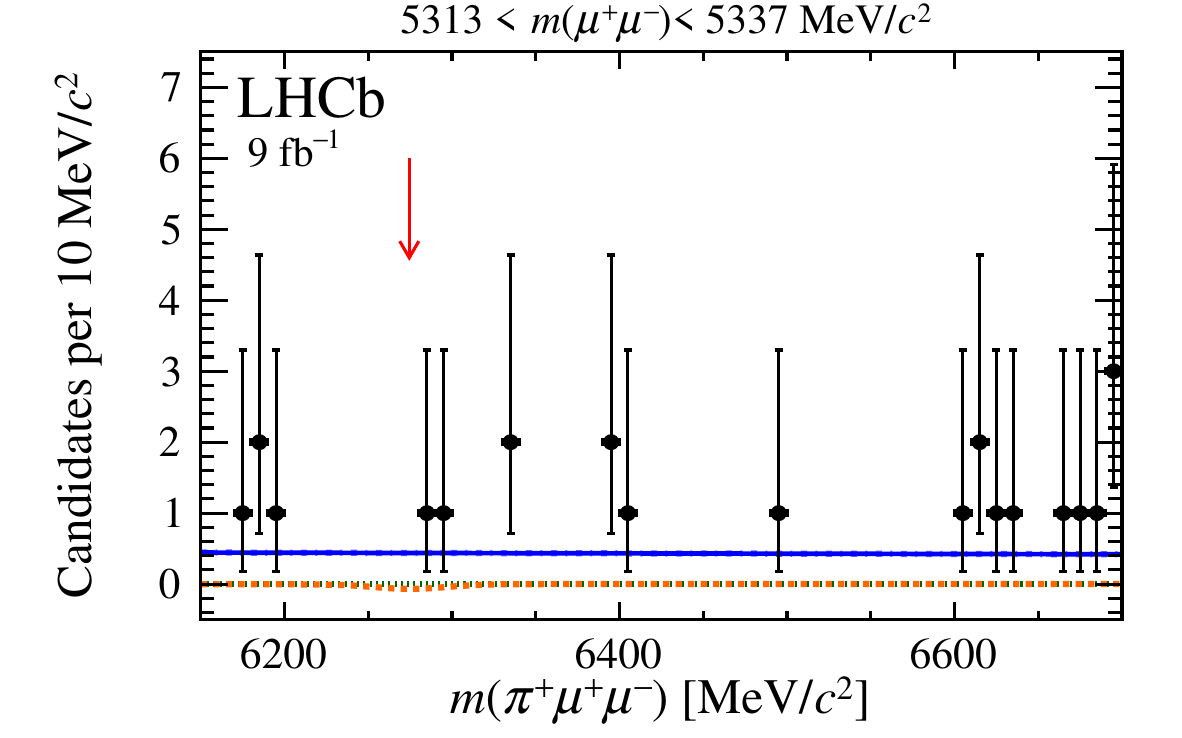}\hfill
    \includegraphics[width=0.50\textwidth]{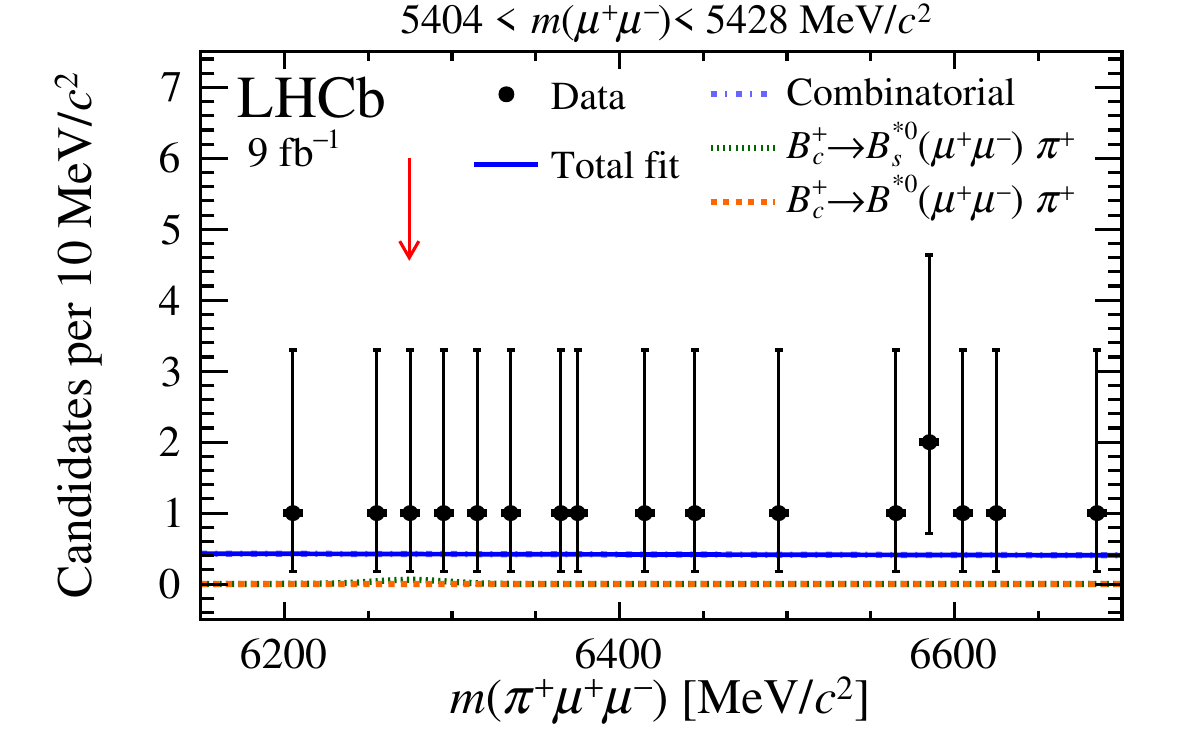}     
    \caption{
    The (top left) two-dimensional distribution of $\pi^+\mu^+\mu^-$ invariant mass versus $\mu^+\mu^-$ invariant mass for the selected \mbox{$B_c^+\to B_{(s)}^{*0}(\mu^+\mu^-)\pi^+$} candidates, together with the (top right) $\mu^+\mu^-$ invariant-mass distribution in the range $6215 < m(\pi^+\mu^+\mu^-)<6335\text{\,Me\kern -0.1em V\!/}c^2$ and the $\pi^+\mu^+\mu^-$ invariant-mass distributions in the ranges (bottom left) $5313 < m(\mu^+\mu^-)<5337\text{\,Me\kern -0.1em V\!/}c^2$ and (bottom right) $5404 < m(\mu^+\mu^-)<5428\text{\,Me\kern -0.1em V\!/}c^2$, with results of the fit superimposed. 
    The areas delimited by the full red lines in the top left figure show the intersections of the aforementioned ranges, which correspond to about \mbox{$\pm3$ times} the experimental resolution around the expected signal-peak positions in each dimension. 
    The red arrows point to the expected signal peak positions.
    The signal yields are consistent with zero and therefore the $B_c^+\to B^{*0}(\mu^+\mu^-)\pi^+$ and $B_c^+\to B_{s}^{*0}(\mu^+\mu^-)\pi^+$ components are barely visible. 
    }
    \label{fig:unblinded-data-2Dscatter}
\end{figure}

Figure~\ref{fig:mass:signal} shows the dimuon and \Bc-candidate invariant-mass distributions of selected \decay{\Bc}{\BorBsstar(\mumu)\pip} candidates, with results of the fit superimposed.
Figure~\ref{fig:unblinded-data-2Dscatter} shows the two-dimensional distribution of selected candidates together with the one-dimensional distributions of dimuon invariant-mass in the range \mbox{$6215 < m(\pip\mumu)<6335\mevcc$} and $\pip\mumu$ invariant-mass in the ranges \mbox{$5313 < m(\mumu)<5337\mevcc$} and \mbox{$5404 < m(\mumu)<5428\mevcc$}.  
These signal-enhanced, one-dimensional distributions are shown with the results of the fit superimposed. 
The yields for the \decay{\Bc}{\Bstarz(\mumu)\pip} and \decay{\Bc}{\Bsstar(\mumu)\pip} decays are consistent with zero.
Table~\ref{tab:fitresults} summarises the yields obtained from the fit.
The fit returns a correlation between the two signal yields of $1.2\%$. 

\clearpage

{\renewcommand{\arraystretch}{1.2}
\begin{table}[t]
    \centering
    \caption{Yields obtained from the fit to data described in the text, with statistical uncertainties only.}    
    \begin{tabular}{cr@{$\,$}l}
       \hline
       Component   & \multicolumn{2}{c}{Yield}\\
       \hline
       \decay{\Bc}{\Bstarz(\mumu)\pip}   & $-0.4$  & $_{-1.1}^{+1.9}$ \\
       \decay{\Bc}{\Bsstar(\mumu)\pip}                                 & $ 0.4  $  & $_{-1.3}^{+2.2}$ \\
       Combinatorial background                                            & $ 282  $  & $\pm\,17  $ \\\hline
    \end{tabular}
    \label{tab:fitresults}
\end{table}
}

\section{Efficiencies and systematic uncertainties} 
\label{sec:Systemtics} 

Table~\ref{tab:syst_constrs} summarises the parameters entering the determination of the single-event sensitivities in Eq.~\eqref{eq:R_BF_norm}, with statistical and systematic uncertainties added in quadrature.
Systematic uncertainties on the signal yields are not included in this procedure, but are instead studied as a cross-check of the procedure and are found to be negligible.

The efficiency ratios between signal and normalisation modes are obtained from simulation accounting for the geometrical acceptance of the detector as well as effects related to the triggering, reconstruction and selection of the \Bc~candidates.
The uncertainties on the efficiency ratios take into account the simulation sample size, uncertainties on the weights applied to the simulation, 
the matching between reconstructed and generated particles in the simulation,
variations of the software trigger requirements,
and the uncertainty on the known \Bc~lifetime. 
All variations are made consistently for the signal and normalisation modes to avoid overestimation of the uncertainty on the efficiency ratio. 

{\renewcommand{\arraystretch}{1.2}
\begin{table}[b]
    \centering
    \caption{Input parameters used in the estimation of the ratio ${\cal R}_{\BorBsstar(\mumu)\pip/\jpsi\pip}$.
    The \mbox{\decay{\jpsi}{\mumu}} branching fraction and its uncertainty are taken from Ref.~\cite{PDG2024}.
    For the $\decay{\Bc}{\jpsi\pip}$ yield the uncertainties are statistical and systematic, respectively; all other uncertainties are systematic.
}  
    \begin{tabular}{c r@{$\,\pm\,$}l} 
    \hline
    Parameter & \multicolumn{2}{c}{Value} \\\hline
       $\BF\left(\decay{\jpsi}{\mumu}\right)$ & $(59.61$ & $0.33)\times 10^{-3}$ \\
       $N_{\jpsi\pip}$ & $6213$ & $89\pm27$\\         
       ${\varepsilon_{\jpsi\pip}}/{\varepsilon_{\Bstarz\pip}}$ & $1.09 $ & $0.05$ \\
       ${\varepsilon_{\jpsi\pip}}/{\varepsilon_{\Bsstar\pip}}$ & $1.18 $ & $0.05$ \\       
       \hline
    \end{tabular}
    \label{tab:syst_constrs}
\end{table}
}

The systematic uncertainties associated with the weights are evaluated by varying all weights within their statistical uncertainties and by varying the binning scheme used to estimate them. 
The systematic uncertainty associated with the multivariate weighting algorithm~(see Sec.~\ref{sec:Detector}) is evaluated by comparing the results obtained with the default and with an alternative algorithm. 
The default algorithm is trained to correct for discrepancies between data and simulation associated with the event track multiplicity and with the transverse momentum and the vertex quality of the \Bc~candidates.
The alternative algorithm is trained using the impact parameter significance of the two muons as additional inputs. 

The systematic uncertainty associated with the matching between reconstructed and generated particles in the simulation is evaluated by comparing the efficiencies obtained including or excluding \B~candidates for which one or more decay products are not correctly matched. 
The systematic uncertainty associated with variations of the software trigger requirements that are not reproduced by the simulation is evaluated by comparing the efficiencies obtained by applying the tightest thresholds and by applying average thresholds within each data-taking period.  
The systematic uncertainty associated with the \Bc~lifetime is evaluated by varying the \Bc\ lifetime in simulation within its uncertainties~\cite{PDG2024}.

A further systematic effect associated with the track reconstruction~\cite{LHCb-DP-2013-002} can arise due to the possible difference in hadronic interactions for the pion tracks in the signal and the normalisation modes (due to the kinematic differences between the decays) and discrepancies between data and simulation in the detector material. This effect is studied in simulation and data and is found to have a negligible impact. 

The effect of the multivariate weighting algorithm has the largest impact on the systematic uncertainty of the efficiency ratio leading to a relative uncertainty around $5\%$. 
The remaining systematic uncertainties cancel out almost completely in the determination of the efficiency ratios and lead to relative uncertainties below $1\%$.  

The normalisation mode yield obtained in the previous section can be affected by the fit model choice and by the assumption of the polarisation of the partially reconstructed backgrounds.
To study the effect of the fit model choice, each fit is performed in three configurations: using the baseline fit model and using two alternative fit models. 
In the two alternative fit models, the analytical function used for the combinatorial background is replaced by a sigmoid function. 
In the first alternative model the same parametrisation as in the nominal model is kept for the other fit components, while in the second alternative model the
\decay{\Bc}{\jpsi\pip} and the misidentified background models are replaced by a modified hyperbolic distribution with power-law tails~\cite{Santos:2013gra} and the model for the partially reconstructed background is replaced by a Gaussian function with a power-law tail to the right side of the distribution.
For the normalisation mode yield, the largest difference between the results obtained with the baseline and alternative models is assigned as systematic uncertainty.
For the global peak shift and width scaling factors, the model choice is found to have a negligible impact. 

In the nominal normalisation mode fit, the \rhop~meson in the \decay{\Bc}{\jpsi\rhop} partially reconstructed background is assumed to be unpolarised. 
However, the polarisation of the \rhop~meson can affect the momentum of the missing pion and hence the \Bc-candidate mass shape of the partially reconstructed backgrounds.
The fit is therefore repeated assuming either full longitudinal or full transverse \rhop~polarisation. 
The difference in the results for the two configurations is found to be negligible.

\section{Results for relative branching fractions} 
\label{sec:Result} 

The fit is repeated with a model identical to that described in Sec.~\ref{sec:Fit}, except that the signal yields are parametrised in terms of branching fraction ratios ${\cal R}_{\BorBsstar(\mumu)\pip/\jpsi\pip}$ using Eq.~\eqref{eq:R_BF_norm}.
The systematic uncertainties associated with the single-event sensitivities are accounted for through Gaussian constraints in the fit.
Using the parameters in Table~\ref{tab:syst_constrs} to calculate the single-event sensitivities gives $\alpha_{\Bstarz\pip}^{\rm SES} = (1.04\pm 0.05)\times 10^{-5}$ and $\alpha_{\Bsstar\pip}^{\rm SES} = (1.13\pm 0.05)\times 10^{-5}$ taking statistical and systematic uncertainties into account.
Including all constraints, the fit yields
\begin{align*}
    {\cal R}_{\Bstarz(\mumu)\pip/\jpsi\pip} &= (-0.44\,_{-1.12}^{+1.99})\times 10^{-5}\,,\\ 
    {\cal R}_{\Bsstar(\mumu)\pip/\jpsi\pip} &= \phantom{-} (0.43\,_{-1.41}^{+2.45})\times 10^{-5}\,.  
\end{align*}
To assess the impact of the systematic uncertainties, the fits are repeated fixing the nuisance parameters to their central values. 
The difference in the uncertainties between the two configurations is around $10^{-7}$, showing that the impact of the systematic uncertainties is negligible.

\begin{figure}[b]
    \centering
    \includegraphics[width=0.5\textwidth]{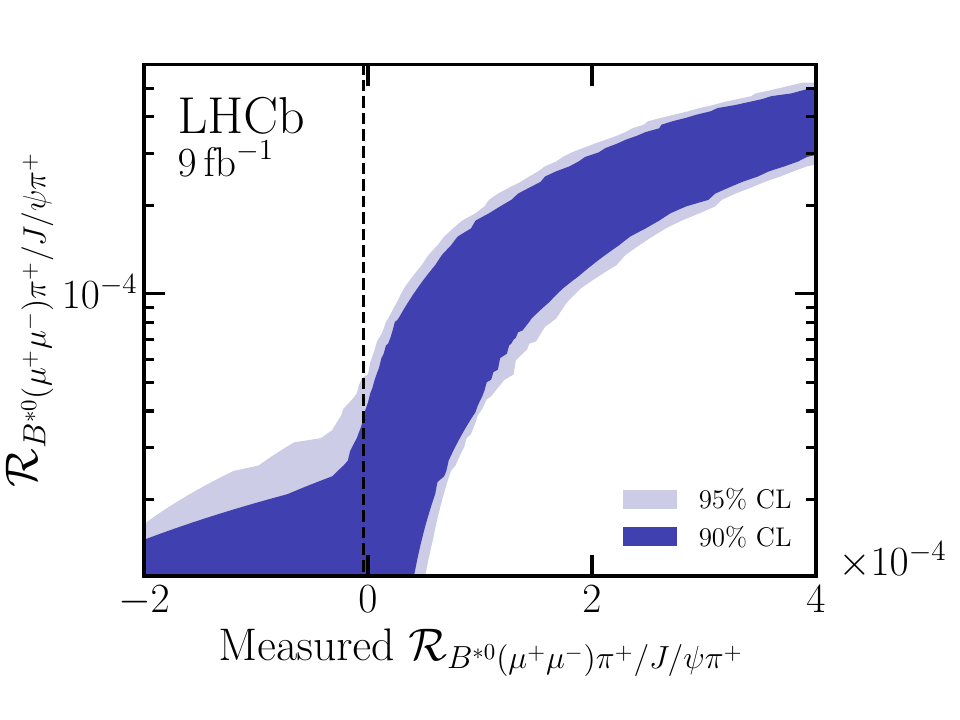}\hfill
    \includegraphics[width=0.5\textwidth]{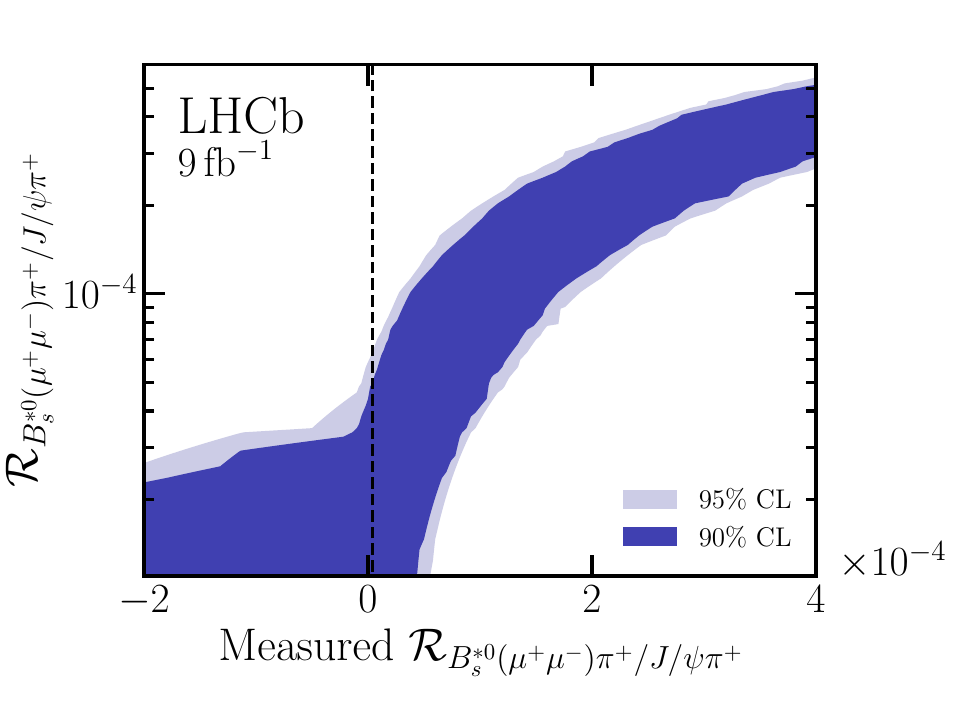}       
    \caption{
    Confidence belts generated using pseudoexperiments according to the Feldman--Cousins prescription.
    The vertical black line shows the results of the fit to data.
    }
    \label{fig:conf_bands}
\end{figure}

Upper limits on the branching fraction ratios are obtained following the \mbox{Feldman--Cousins} prescription~\cite{Feldman:1997qc}: pseudoexperiments are generated for various values of ${\cal R}_{\BorBsstar(\mumu)\pip/\jpsi\pip}$ and the distributions of the measured ${\cal R}_{\BorBsstar(\mumu)\pip/\jpsi\pip}$ values in the pseudoexperiments are used to form confidence belts. 
Nuisance parameters are varied within their uncertainties in the generation of the pseudoexperiments.
The scan to obtain limits for ${\cal R}_{\Bstarz(\mumu)\pip/\jpsi\pip}$ is performed assuming that ${\cal R}_{\Bsstar(\mumu)\pip/\jpsi\pip}$ is zero and vice versa. 
This assumption does not impact the obtained limits as the correlation between the signal yields is negligible.
Figure~\ref{fig:conf_bands} shows confidence belts at $90\%$ and $95\%$\,CL. 
Evaluation of the upper limits at the central values obtained from the fit to data yields
\begin{align*}
    {\cal R}_{\Bstarz(\mumu)\pip/\jpsi\pip} &<  3.8\,(5.2)\times 10^{-5}  \text{ at 90 (95)\%\,CL}\,,\\ 
    {\cal R}_{\Bsstar(\mumu)\pip/\jpsi\pip} &< 5.0\,(6.3)\times 10^{-5}  \text{ at 90 (95)\%\,CL}\,.  
\end{align*}
As further checks, the procedure is repeated restricting the signal yield to positive values,  
or replacing in the fit model the signal parametrisation with the sum of two Gaussian functions, one with power-law tails.
No significant changes in the obtained upper limits are found.

As a further cross-check, the ratio $\BF(\decay{\Bc}{\psitwos(\mumu)\pip})/\BF(\decay{\Bc}{\jpsi(\mumu)\pip})$ is measured following the same analysis procedure as in Ref.~\cite{LHCb-PAPER-2023-037}, but applying the BDT classifier, $\cos{\theta_l}$ and particle-identification requirements optimised for this work.
The measured value is $0.279\pm 0.025$, where the uncertainty is statistical only. This is consistent within $1.5$ standard deviations with the results in Ref.~\cite{LHCb-PAPER-2023-037} when accounting for correlations, and it also agrees with previously published measurements of this quantity~\cite{LHCb-PAPER-2012-054,LHCb-PAPER-2015-024}.

\section{Summary} 
\label{sec:Summary} 

A search is performed for the very rare \decay{\Bstarz}{\mumu} and \decay{\Bsstar}{\mumu} decays by analysing \mbox{\decay{\Bc}{\pip\mumu}} decays. 
The analysis uses proton-proton collision data collected with the LHCb detector between 2011 and 2018, corresponding to an integrated luminosity of 9\invfb.
No evidence for an excess of signal events over background is observed for the two decay modes and the first upper limits on their branching fraction ratios are obtained,
\begin{align*}
    {\cal R}_{\Bstarz(\mumu)\pip/\jpsi\pip} &< 3.8\times 10^{-5}\,,\\ 
    {\cal R}_{\Bsstar(\mumu)\pip/\jpsi\pip} &< 5.0\times 10^{-5}\,, 
\end{align*}
at $90\%$ confidence level.
Currently, there are no measurements available for the ratio \mbox{$\BF(\decay{\Bc}{\BdorBsstar\pip})/\BF(\decay{\Bc}{\jpsi\pip})$}. 
The branching fraction for the $\decay{\Bc}{\Bsstar\pip}$ decay is expected to be of the same order of magnitude as that for the \mbox{$\decay{\Bc}{\Bs\pip}$}~decay~\cite{Chang:1992pt,Anisimov:1998uk,Colangelo:1999zn,AbdEl-Hady:1999jux,Kiselev:2002vz,Ebert:2003wc,Sun:2008wa,Dhir:2008hh,Ivanov:2006ni,Likhoded:2010jr,Naimuddin:2012dy,Shi:2016gqt}, and therefore the ratio $\BF(\decay{\Bc}{\Bsstar\pip})/\BF(\decay{\Bc}{\jpsi\pip})$ is expected to be of the order $\mathcal{O}(10^2)$~\cite{LHCb-PAPER-2022-034}.
Based on this assumption, the result for ${\cal R}_{\Bsstar(\mumu)\pip/\jpsi\pip}$ is expected to correspond to an upper limit on the absolute branching fraction \mbox{$\BF(\decay{\Bc}{\Bsstar\pip})$} of the order $\mathcal{O}(10^{-6})$.
There are no published results on the branching fractions \mbox{$\BF(\decay{\Bc}{\BdorBdstar\pip})$}, but these are expected to be Cabibbo-suppressed relative to those for \decay{\Bc}{\BsorBsstar\pip} decays, leading to expectations for the upper limits that are correspondingly higher.
Once measurements of the ratio $\BF(\decay{\Bc}{\BdorBsstar\pip})/\BF(\decay{\Bc}{\jpsi\pip})$ become available, it will be possible to translate the results in this paper into upper limits on the absolute \decay{\BdorBsstar}{\mumu} branching fractions.

\section*{Acknowledgements}
%
%
\noindent We express our gratitude to our colleagues in the CERN
accelerator departments for the excellent performance of the LHC. We
thank the technical and administrative staff at the LHCb
institutes.
We acknowledge support from CERN and from the national agencies:
CAPES, CNPq, FAPERJ and FINEP (Brazil); 
MOST and NSFC (China); 
CNRS/IN2P3 (France); 
BMBF, DFG and MPG (Germany); 
INFN (Italy); 
NWO (Netherlands); 
MNiSW and NCN (Poland); 
MCID/IFA (Romania); 
MICIU and AEI (Spain);
SNSF and SER (Switzerland); 
NASU (Ukraine); 
STFC (United Kingdom); 
DOE NP and NSF (USA).
We acknowledge the computing resources that are provided by CERN, IN2P3
(France), KIT and DESY (Germany), INFN (Italy), SURF (Netherlands),
PIC (Spain), GridPP (United Kingdom), 
CSCS (Switzerland), IFIN-HH (Romania), CBPF (Brazil),
and Polish WLCG (Poland).
We are indebted to the communities behind the multiple open-source
software packages on which we depend.
Individual groups or members have received support from
ARC and ARDC (Australia);
Key Research Program of Frontier Sciences of CAS, CAS PIFI, CAS CCEPP, 
Fundamental Research Funds for the Central Universities, 
and Sci. \& Tech. Program of Guangzhou (China);
Minciencias (Colombia);
EPLANET, Marie Sk\l{}odowska-Curie Actions, ERC and NextGenerationEU (European Union);
A*MIDEX, ANR, IPhU and Labex P2IO, and R\'{e}gion Auvergne-Rh\^{o}ne-Alpes (France);
AvH Foundation (Germany);
ICSC (Italy); 
Severo Ochoa and Mar\'ia de Maeztu Units of Excellence, GVA, XuntaGal, GENCAT, InTalent-Inditex and Prog. ~Atracci\'on Talento CM (Spain);
SRC (Sweden);
the Leverhulme Trust, the Royal Society
 and UKRI (United Kingdom).

\addcontentsline{toc}{section}{References}
\bibliographystyle{LHCb}
\bibliography{main,standard,LHCb-PAPER,LHCb-CONF,LHCb-DP,LHCb-TDR}

\newpage
\centerline
{\large\bf LHCb collaboration}
\begin
{flushleft}
\small
R.~Aaij$^{36}$\lhcborcid{0000-0003-0533-1952},
A.S.W.~Abdelmotteleb$^{55}$\lhcborcid{0000-0001-7905-0542},
C.~Abellan~Beteta$^{49}$,
F.~Abudin{\'e}n$^{55}$\lhcborcid{0000-0002-6737-3528},
T.~Ackernley$^{59}$\lhcborcid{0000-0002-5951-3498},
A. A.~Adefisoye$^{67}$\lhcborcid{0000-0003-2448-1550},
B.~Adeva$^{45}$\lhcborcid{0000-0001-9756-3712},
M.~Adinolfi$^{53}$\lhcborcid{0000-0002-1326-1264},
P.~Adlarson$^{80}$\lhcborcid{0000-0001-6280-3851},
C.~Agapopoulou$^{13}$\lhcborcid{0000-0002-2368-0147}
% ...

{\footnotesize \it

% ...
$^{13}$Universit{\'e} Paris-Saclay, CNRS/IN2P3, IJCLab, Orsay, France\\
% ...
$^{36}$Nikhef National Institute for Subatomic Physics, Amsterdam, Netherlands\\
% ...
$^{45}$Instituto Galego de F{\'\i}sica de Altas Enerx{\'\i}as (IGFAE), Universidade de Santiago de Compostela, Santiago de Compostela, Spain\\
% ...
$^{49}$Physik-Institut, Universit{\"a}t Z{\"u}rich, Z{\"u}rich, Switzerland\\
% ...
$^{53}$H.H. Wills Physics Laboratory, University of Bristol, Bristol, United Kingdom\\
% ...
$^{55}$Department of Physics, University of Warwick, Coventry, United Kingdom\\
% ...
$^{58}$School of Physics and Astronomy, University of Glasgow, Glasgow, United Kingdom\\
$^{59}$Oliver Lodge Laboratory, University of Liverpool, Liverpool, United Kingdom\\
% ...
$^{67}$Syracuse University, Syracuse, NY, United States\\
% ...
$^{80}$Department of Physics and Astronomy, Uppsala University, Uppsala, Sweden, associated to $^{58}$\\
% ...
}
\end{flushleft}

\end{document}